\documentclass[aps,twocolumn,groupedaddress]{revtex4}
\usepackage{graphics}
\usepackage{bm}
\usepackage{amssymb}
\usepackage{stmaryrd}
\usepackage[usenames]{color}

\begin{document}
\title{Loose packings of frictional spheres}
\author{Greg R. Farrell\footnote{E-mail: \textit{grfarrell@physics.umass.edu}},
 		K. Michael Martini,
		Narayanan Menon\footnote{E-mail: \textit{menon@physics.umass.edu}}	
}
\affiliation{Department of Physics, University of Massachusetts, Amherst MA 01003, U.S.A.}

\begin{abstract}
We have produced loose packings of cohesionless, frictional spheres by sequential deposition of highly-spherical, monodisperse
particles through a fluid. By varying the properties of the fluid and the particles, we have identified the Stokes number ($St$) -- rather than the buoyancy of
the particles in the fluid -- as the parameter controlling the approach to the loose packing limit. The loose packing limit is attained at a threshold value of $St$ at which the
kinetic energy of a particle impinging on the packing is fully dissipated by the fluid. Thus, for cohesionless particles,  the dynamics of the deposition
process, rather than the stability of the static packing, defines
the random loose packing limit.
We have made direct measurements of the interparticle friction in the fluid, and
present an experimental measurement of the loose packing volume fraction, $\phi_{RLP}$,
as a function of the friction coefficient $\mu_s$.
\end{abstract}

\maketitle
\section{Introduction}

The most elementary characteristic of a disordered sphere packing is the fraction, $\phi$,
of the total volume occupied by particles. Stable packings of cohesionless, frictional, spheres
exist over a broad range of volume fractions\cite{Scott1960,Macrae1961,Rutgers1962,Scott1969}.
The term ``random close packed'' (RCP) refers to the upper bound $\phi_{RCP}$ on the volume fraction
at which a packing of identical spheres can be prepared without introducing crystalline ordering. Packings
of  $\phi \approx 0.64$, are consistently achieved in experiments and simulations; this number is
insensitive to variations in interparticle forces and to the compaction protocol, however,
questions remain as to whether there exists a tight upper bound \cite{Chaudhuri2010}.
The robustness of $\phi_{RCP}$ has motivated attempts to understand RCP in purely geometric
terms\cite{Jalali2004} or in terms of the statistical mechanics of hard spheres or soft spheres at zero
temperature\cite{Torquato2000,Gao2006,Kamien2007}. In this article we experimentally explore
the \textit{lower} bound on volume fractions of mechanically stable packings of frictional,
noncohesive, identical, hard spheres. Specifically, our questions are: does a loose packing limit exist for frictional but cohesionless spheres? Will this loose
packing limit depend on the properties of the particle and the
preparation protocol or, like the RCP limit, will it be robust to changes
in these variables, and possibly admit descriptions in terms of the
statistical mechanics of hard spheres\cite{Bernal1964}?

In the first systematic study of loose packings, \citet*{Onoda1990} sedimented glass spheres in fluids of varying densities $\rho_{f}$, approaching the density $\rho_{s}$ of the
sphere. They found that the packing fraction approached an
asymptotic ``random loose packed'' (RLP) value,
$\phi_{RLP}=0.555$, in the limit of vanishing gravitational
acceleration in the fluid, $g_{f}\equiv g\,(1-\rho_{f}/\rho_{s}) \rightarrow 0$. 
However, the limit $g_{f}\rightarrow 0$ conflates two different physical effects, 
both of which may plausibly lead to lower volume fractions.
The first effect involves the dynamics of assembling the packing: as
$g_{f} \rightarrow 0$, falling spheres reach the packing with less
inertia to explore the surface and rearrange their neighbors. This
can trap the particles in higher-energy, fluffier packings.
A second, distinct, effect concerns the statics of the structure: as neutral buoyancy is approached, more fragile packings may become stable since the
gravitational load borne by the packing vanishes relative to weak
cohesive forces.  Indeed, simulations by \citet{Dong2006} argue that 
attractive van der Waals forces are important in stabilizing the 
packings of \citeauthor{Onoda1990} at small $g_{f}$.
Arbitrarily low packing fractions can be attained when attractive interparticle forces are dominant\cite{Weitz1984}, which calls into question the ability to experimentally access an RLP limit for cohesionless spheres.

A key goal of our experiments is to peel apart the distinct effects of fluid properties on the statics and dynamics.
Independently controlling the viscosity and density of the fluid allows us to test whether a unique RLP limit is reached as gentle deposition conditions are approached along arbitrary directions in the density--viscosity plane.
Approaching the limit of gentle deposition by increasing viscosity while keeping the gravitational load finite allows us to avoid the cohesive regime and test whether a well-defined $\phi_{RLP}$ exists for noncohesive spheres.

In the absence of friction, the RLP and the RCP limits are believed
to coincide\cite{OHern2002}. However, as discussed above, stable 
packings with $\phi<\phi_{RCP}$ are common, with the packing fraction 
showing some material-dependence \cite{Scott1960,Rutgers1962}.  
The relevant material property has been conjectured to be surface
roughness \cite{Scott1969,Onoda1990} and is experimentally found 
to correlate with angle of repose\cite{Jerkins2008}. Thus, the cause 
of this variability is generally modelled as a friction coefficient.  
Simulations with friction find the RLP limit to be a systematically 
decreasing function of friction \cite{Zhang2001,Silbert2002,Song2008,Ciamarra2008} 
albeit with unexpectedly large values ($\mu\approx1.$) needed to 
reproduce values seen experimentally\cite{Onoda1990,Jerkins2008}. 
In this work we directly explore $\phi_{RLP}$ as a function of measured
friction coefficient with noncohesive spheres. We emphasize that
particle contacts can exert normal and tangential forces, but cannot 
support tension.

Another goal of our experiments, complementary to previous work, is
to produce packings by the sequential addition of particles, rather
than by collective procedures.
Recent experiments by \citeauthor{Jerkins2008} have studied settling following brief pulses of flow in a liquid-fluidized bed\cite{Jerkins2008} arriving at volume fractions similar to those produced by collective sedimentation\cite{Onoda1990}.  Simulations of sedimented packings\cite{Zhang2001,Silbert2002} have studied volume fraction as a function of particle
properties such as friction and inelasticity.
Other simulations\cite{OHern2002,Song2008} generate disordered packings by collectively relaxing particle configurations
as volume, pressure, temperature or particle interactions are smoothly varied.
However, packings created by sequential deposition, in which a particle
comes to rest at the first mechanically stable location that it
encounters, may lead to different packings than those obtained by
collective preparation protocols.

In this article we present data on loose sphere packings
prepared by the sequential sedimentation of
frictional, non-cohesive spheres large enough to eliminate the
influence of van der Waals and other attractive forces.  By using
fluids of varying density and viscosity, we identify the parameter
in the deposition process that controls passage to a putative
RLP limit. Rather than $g_{f}$, this parameter is the 
Stokes number, $St\equiv(2/9)\,\rho_s\,V\,r/\eta$, where $r$ 
is the radius of the sphere, $V$ its velocity and $\eta$
is the dynamic fluid viscosity.  We also vary the friction between
spheres, both by varying the material and by increasing surface
roughness via controlled etching. We find that the packing
fraction in the loose packing limit is a function of interparticle
friction, the values of which are quite high, a result in
qualitative agreement and simulational findings.

\section{Experimental Methods}

\begin{figure}[ht]
\scalebox{1.0}{\includegraphics{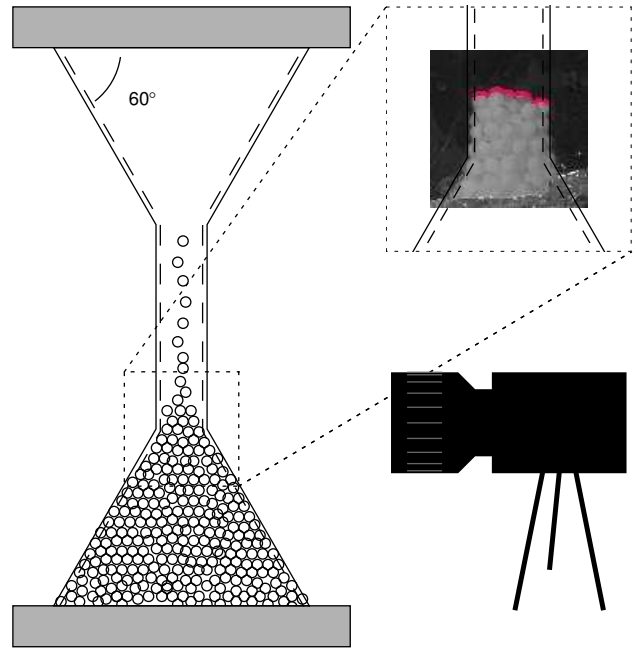}} \caption{Hour-glass shaped
sedimentation apparatus. Top right: Image of the topography of the
top surface of the particles.  Since the volume is estimated from a
single projection, there is a small systematic positive bias
of $\delta\phi \lesssim +0.002$ in measuring the volume of the packing.}
\label{apparatus}
\end{figure}

\begin{figure}[ht]
\scalebox{1.0}{\includegraphics{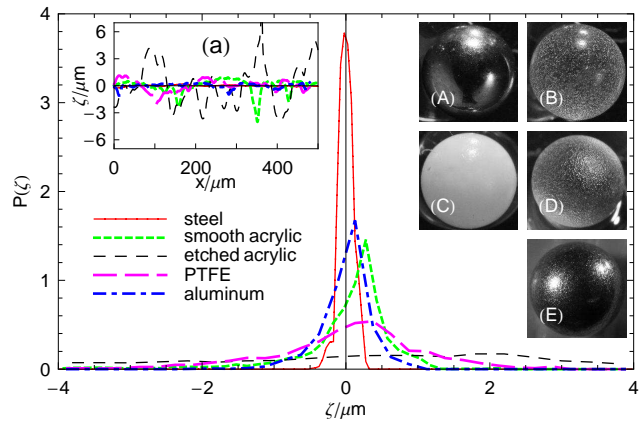}}
\caption{Histograms of roughness $\zeta$ of spherical particles as
measured with a Dektak3 profilometer.  Inset (a): Select profilometer traces $\zeta$ as a function of distance along the surface $x$ (the gross curvature of the sphere has been subtracted).  Images of spherical particles: (A) steel, (B) smooth acrylic, (C) PTFE, (D) solvent-etched acrylic, (E) aluminum.}
\label{roughHist}
\end{figure}

%This ensures that colors match those in the figures.
\definecolor{001MNa}{rgb}{0., 0.981008, 0.836745}
\definecolor{476524}{rgb}{0.682269, 0.807899, 0.}
\definecolor{5050do}{rgb}{0.801982, 0.619908, 0.}
\definecolor{709291}{rgb}{0.829836, 0.322324, 0.}
\definecolor{dodeca}{rgb}{0., 0., 1.}
\definecolor{Fluka0}{rgb}{0., 0.219067, 0.998808}
\definecolor{heavym}{rgb}{0., 0.965987, 0.573886}
\definecolor{heptan}{rgb}{0.964119, 0., 0.740763}
\definecolor{highte}{rgb}{0., 0.97316, 0.684894}
\definecolor{lightm}{rgb}{0.0709943, 0.894864, 0.}
\definecolor{pentan}{rgb}{0.878014, 0., 0.983364}
\definecolor{water}{rgb}{0.886786, 0., 0.053317}
\begin{table*}%[h]
\caption{Fluids used in hour-glass sedimentation experiments.  The symbols match those used in all figures.  Mixture ratios reflect contents before degassing.  Density and viscosity were measured after degassing.  ${}^{*}$lit.\cite{CRC}.  ${}^{\dagger}$our measurement.  ${}^{\ddagger}$lit. at 20C\cite{CRC}.  ${}^{\S}$MSDS.}
\begin{tabular}{|l|c|c|c|}\hline
fluid&                                                              density/$(g\,cm^{-3})$&         viscosity/$(mPa\,s)$&   symbol                                                              \\ \hline\hline
n-pentane (Fisher Sci)&                                                 $0.626^{\ddagger}$&   $0.23^{*}$&                           \color{pentan}{$\blacktriangleleft$}    \\ \hline
n-heptane (Fisher Sci)&                                                 $0.684^{\ddagger}$&   $0.41^{*}$&                           \color{heptan}{$\triangleright$}            \\ \hline
water (Millipore)&                                                      $1.00^{\ddagger}$&    $1.03^{*}$&                           \color{water}{$\triangledown$}              \\ \hline
n-dodecane (Fisher Sci)&                                                $0.75$&                           $1.53^{*}$&                           \color{dodeca}{$\blacktriangle$}            \\ \hline
$\sim$71/29 n-dodecane/light mineral oil mixture&   $0.78^{\dagger}$& $3.02^{\dagger}$&             \color{709291}{$\circ$}                             \\ \hline
$\sim$50/50 n-dodecane/light mineral oil mixture&   $0.80^{\dagger}$& $5.96^{\dagger}$&             \color{5050do}{$\boxempty$}                     \\ \hline
$\sim$48/52 n-dodecane/light mineral oil mixture&   $0.82^{\dagger}$& $11.5^{\dagger}$&             \color{476524}{$\blacklozenge$}             \\ \hline
light mineral oil (Fisher Sci)&                                 $0.83^{\S}$&              $46.0^{\dagger}$&             \color{lightm}{$\lozenge$}                      \\ \hline
high temperature silicone oil (Acros Organics)& 		$1.05^{\S}$&              $117.^{\dagger}$&             \color{highte}{$\bullet$}                           \\ \hline
0.01M NaCl in $\sim$70/30 propylene glycol/glycerol&			$1.09^{\dagger}$&	$125.^{\dagger}$&							 \color{001MNa}{$\blacktriangleright$}	                           \\ \hline
heavy mineral oil (Fisher Sci)&                                 $0.83^{\S}$&              $157.^{\dagger}$&             \color{heavym}{$\triangleleft$}             \\ \hline
Fluka 08577 Density and Viscosity Standard&         $0.87^{\S}$&              $1270.^{\S}$&                     \color{Fluka0}{$\blacktriangledown$}    \\ \hline
\end{tabular}
\label{fluidtable}
\end{table*}

\begin{table*}%[h]
\caption{Properties of sets of spheres.  We measured particle diameter using a technique similar to Scott's\cite{Scott1969}, measuring
the length of $\sim 100$ spheres in a groove.  Density was calculated from
this diameter and the weight of these spheres.  To quantify
polydispersity and sphericity, we measured the diameter of
individual spheres with a machinist's micrometer accurate to $2.5\mu
m$, along five or more directions.  ``Asphericity'' is the relative deviation from sphericity calculated as the standard deviation of these diameter measurements relative to the mean.
``Polydispersity'' is the standard deviation of the average
diameters of a set of 20 spheres relative to the mean. ``RMS roughness'' gives the root-mean-square deviation from sphericity of profilometer traces taken of the sphere's surface(see Fig.~\ref{roughHist}).
$\bar{\mu_s}$ and $\sigma_{\mu_s}$ are the mean and width of the
distribution of static friction coefficient of gently contacting
spheres in fluid (see Fig.~\ref{frictionFit}).}
\begin{tabular}{|l|c|c|c|c|c|c|c|}\hline
sphere material& diameter/$cm$&          density/$(g\,cm^{-3})$& 		polydispersity&    	asphericity&         RMS roughness/$\mu m$&  $\bar{\mu_s}$&  	 $\sigma_{\mu_s}$\\ \hline\hline
PTFE&           $0.3205(4)$&     $2.1389(9)$&         $0.21\pm0.06\%$&    $0.14\pm0.06\%$&    $1.1\pm0.6$&       			 $0.540\pm0.003$&	0.10    \\  \hline
aluminum&	  		$0.3191(3)$&  		$2.775(3)$&						$\leq 0.04\%$&			$\leq 0.04\%$&			$0.32\pm0.14$&					 $0.62\pm0.06$&		0.16		\\  \hline
steel&          $0.3179(4)$&     $7.774(7)$&           $0.14\pm0.05\%$&    $\leq 0.04\%$&   		$0.10\pm0.02$&          $0.66\pm0.14$&		 0.15    \\  \hline
smooth acrylic& $0.3174(3)$&     $1.1800(9)$&         $0.15\pm0.06\%$&    $0.06\pm0.04\%$&  	$0.7\pm0.3$&       			$0.88\pm0.03$&		 0.10   \\  \hline
etched acrylic& $0.3092(3)$&     $1.1741(9)$&         $0.16\pm0.07\%$&   	$0.08\pm0.05\%$&    $2.6\pm0.1$&            $0.96\pm0.03$&		 0.10    \\  \hline
\end{tabular}
\label{beadtable}
\end{table*}

We prepared mechanically stable packings of monodisperse spheres
immersed in fluids, in an hour-glass shaped apparatus
(Fig.~\ref{apparatus}) placed on a vibration-isolation table. Using a
variety of fluids (Table~\ref{fluidtable}) and spheres (Table~\ref{beadtable})
we formed packings under a wide range of viscosity and buoyancy conditions
allowing distinctions to be drawn between the relative merits of
different parameters controlling approach to the RLP limit.

Packings were prepared by inverting the hour-glass shaped cell and
allowing particles to settle through the fluid under gravity. The
hour-glass geometry consists of two conical sections with a cone
angle of $60\,^{\circ}$ and base diameter of 24 sphere-diameters ($d=2\,r$)
connected by a cylindrical neck of $4.2\,d$ in diameter, which is
only as wide as necessary to avoid jamming by arch formation in the
neck. The narrow neck allows the passage of only a few particles at
a time. (We have also deposited spheres singly by a mechanical
dropper, with very similar results to those reported here.) The
packing grows as a conical pile at the angle of repose ($\approx 23-26\,^{\circ}$), which is much smaller than the cone angle of the
container, thus eliminating any empty pockets near the walls. The
conical cell was chosen to minimize the weight supported by the
sloping walls of the container.  No crystalline order was observed
near the bottom or side walls.

Data for the volume fraction of the packing are taken when the top
surface of the packing just enters the neck of the cell. The total
volume of the packing is the volume of the cone plus the small
contribution from the spheres that are in the neck. The latter is
determined from an image of the topography of the top surface of the
particles (inset Fig.~\ref{apparatus}). All volume fractions
reported in this article are subject to the same systematic error in
the range $\delta\phi\approx0.000\text{--}0.002$ due to
uncertainties in these volumes and in the volume of the hour glass.

The particles used in our experiments are commercially available
spheres of acrylic (PMMA), teflon (PTFE), steel, and
aluminum with nominal diameter $d \approx 3.18\,mm$(see
Table~\ref{beadtable}, Fig.~\ref{roughHist}). We also use acrylic spheres which were roughened by timed etches in
acetone. All sets of spheres are very monodisperse and highly
spherical with, at worst a deviation from sphericity of $\sim
10^{-3} d$, a surface roughness of the same order, and a
polydispersity of double this magnitude.  These spheres are much
larger than those used by Onoda and Liniger ($0.25\pm0.02\,mm$,
glass). The advantages of using large spheres are evident: attractive van der
Waals forces are negligibly small compared to other forces in the
problem and the particles are well characterized and of extremely
high sphericity and monodispersity. Thus these experiments represent a
better approximation to the idealized packing of hard, monodisperse
spheres than previous experiments.

\begin{figure*}[ht]
\scalebox{1.0}{\includegraphics{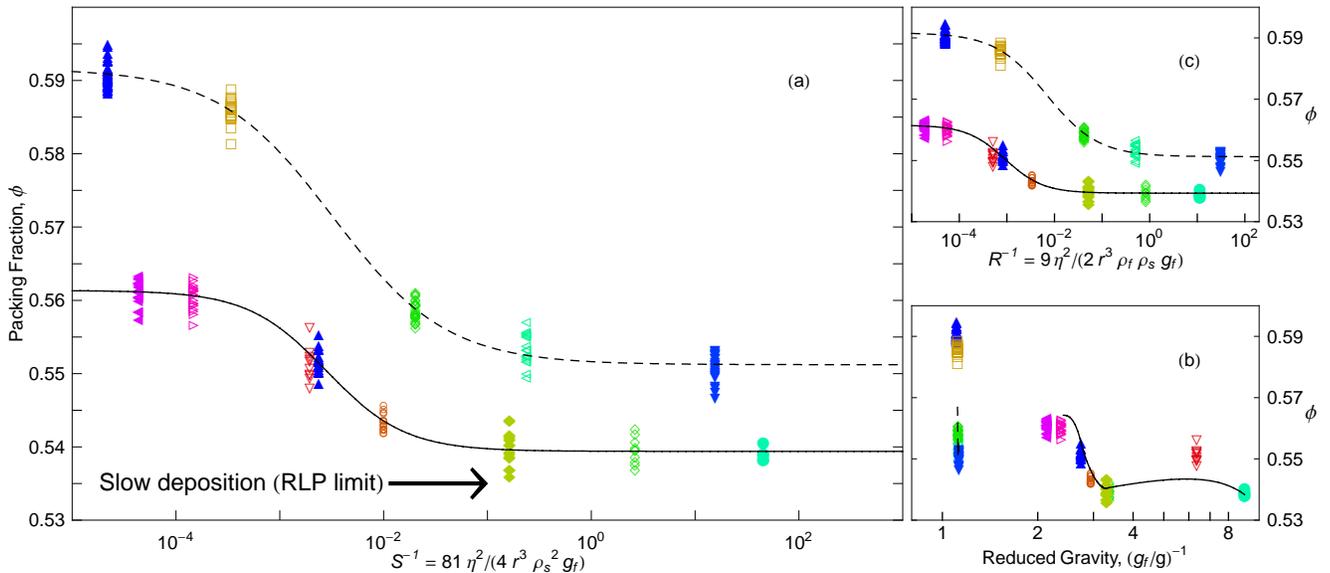}}
\caption{Packing fraction $\phi$ vs. the dimensionless parameters $S$, $R$, and
$g_f/g$. The data points represent individual packings; the spread
in the data is much larger than the random error on each data point.
For a particle falling at low Reynolds number, $S$ is the Stokes
number and $R$ is the Reynolds number itself. $g_{f}$ is the
buoyancy-reduced gravitational acceleration felt by a particle in
the fluid. (Onoda and Liniger's $\Delta g \equiv g_f/g$\cite{Onoda1990}.) In all three graphs the solid line connects data for smooth acrylic spheres and the dashed line is for steel spheres.}
\label{phivsparamsFit}
\end{figure*}

Apart from van der Waals attraction\cite{Dong2006}, experimental
results on the low packing fraction limit are sensitive to other
attractive forces of electrostatic and capillary origin. In setting
up the apparatus, great care was taken to degas fluids and to
introduce the fluid to the spheres slowly to avoid entraining air
bubbles which form attractive bridges between between poorly wetted
surfaces, especially rough ones. Pumping a vacuum on heated, stirred
fluids for hours was often insufficient to avoid the appearance of
attractive forces between spheres in non-wetting fluids, a phenomenon
that has recently been associated with the existence of
long-lived\cite{Attard2002,Borkent2007} nanobubbles capable of
exerting forces comparable to gravity for PMMA spheres in our more
closely density matched fluids\cite{Parker1994}.  To avoid this
phenomenon, we have used well-wetting fluids when possible and
avoided close density matches in poorly-wetting fluids so that the
contribution of attractive forces is negligible. Where charging
effects were suspected, we repeated measurements with salts added to
screen coulomb interactions.

\section{Results and discussion}

The problem of a sphere falling in the fluid involves five
dimensionful parameters: $r=d/2$ and $\rho_{s}$, the radius and
density of the spheres; $\rho_{f}$ and $\eta$, the density and
dynamic viscosity of the fluid; and $g_f=g\,(1-\rho_{f}/\rho_{s})$,
the buoyancy-reduced gravitational acceleration in the fluid. Apart
from $g_f/g$ (as suggested by Ref.~\cite{Onoda1990}), other
pertinent dimensionless groups are the Reynolds number
$Re=2\,\rho_{f}\,V\,r/\eta$ and the Stokes number $St=(2/9)\rho_{s}\,V\,r/\eta$,
where $V$ is the velocity of the particles as they approach
the packing.  In reducing our data we use for $V$, the terminal velocity of
the particle under Stokes drag, leading to the dimensionless
parameters $R=(2/9)\,r^3\,\rho_s\,\rho_f\,g_f/\eta^2$
and $S=(4/81)\,r^3\,\rho_s^2\,g_f/\eta^2$. In the limit
of gentle deposition, $R=Re$, $S=St$.  $S$ can be
interpreted either as a dimensionless damping length, or as the ratio of kinetic energy to the potential energy, $mass \times g_f\,d$, which quantifies
the degree to which a falling sphere can rearrange the packing.

\begin{figure}[ht]
\scalebox{1.0}{\includegraphics{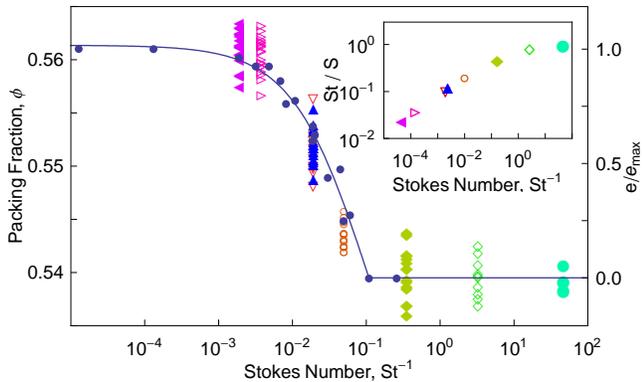}}
\caption{The main plot shows the packing fraction, $\phi$, of acrylic spheres (left vertical axis, symbols match Fig.~\ref{phivsparamsFit}) versus Stokes number, $St$.  We compare the trend with the data of \citeauthor{Gondret2002}(from Fig. 6, Ref.~\cite{Gondret2002}), plotted here in dark circles against the right vertical axis) for the $St$-dependence of the restitution coefficient $e$, scaled by its maximal value $e_{max}$.  The data are for collisions of a teflon sphere in a fluid, however, data for different materials collapse on the same curve.  To make the comparison with Ref.~\cite{Gondret2002}, we inferred the value of $St$ in our experiments from measured values (Table~\ref{beadtable}) and a standard drag curve\cite{Majumder2004}.  This correction is shown in the inset; as expected the correction is small when $St$ is small.}
\label{restCoeff}
\end{figure}

\begin{figure}[ht]
\scalebox{1.0}{\includegraphics{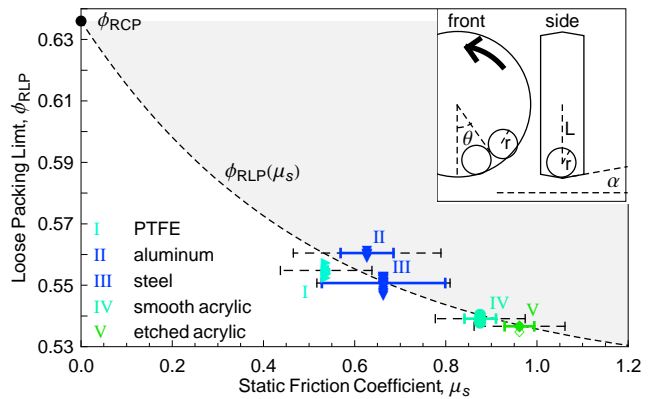}}
\caption{Limiting low packing fraction $\phi_{RLP}$ plotted versus
$\mu_s$, the mean static friction coefficient.  $\mu_s$ is measured
by a new technique where the friction
between spheres immersed in a fluid is ascertained from the maximum
angle at which pairs of spheres in a vertical, circular track
maintain contact due to static friction. The dashed error bars
indicate the width in the distribution of $\mu_s$, and the solid
error bars are uncertainty in the mean of $\mu_s$. The shaded area represents the region of stable, disordered packings. The lower bound (dashed curve) is a guide to the eye.  The inset shows a drawing of the geometry used in the friction measurement (scale: bead radius $r=1.59\,mm$).}
\label{frictionFit}
\end{figure}

We display in Fig.~\ref{phivsparamsFit} the major result of this
paper.  For spheres of a given material, the packing fraction
decreases as sedimentation is done more gently. When plotted against
the dimensionless group $S$, the volume fraction approaches an
asymptotic limiting low volume fraction which we may interpret as
$\phi_{RLP}$. We note the the limiting
$\phi_{RLP}$ is different for spheres of different materials.
The asymptotic limit is directly available from the
data, and unlike in previous measurements need not be obtained by
extrapolation (for which there is no reliable theoretical
guideline). Each data point is taken with a different liquid, and not with a chemical
series or a dilution series; the smooth approach to this limit thus
suggests that the macroscopic parameters of the fluid and sphere are
sufficient to fully characterize the preparation process, and that
any microscopic interparticle interactions mediated by the fluid
have been successfully suppressed.

Furthermore, Fig.~\ref{phivsparamsFit}(b) shows that the degree of
density matching, quantified by $g_{f}/g = 1-\rho_f/\rho_s$, is not an
appropriate parameter to define the RLP limit of this packing
protocol.  Low packing fractions can be achieved without density
matching, by sedimenting in sufficiently viscous fluids. For a fluid
of a given viscosity, RLP can of course be approached by varying
$g_{f}/g$, as the work of Onoda and Liniger suggests. The role of
$g_{f}/g$ is however, not in reducing the static load on the packing
structure, but purely in slowing down the dynamics of the packing
process.

Fig.~\ref{phivsparamsFit}(c) shows that the volume fraction $\phi$
also smoothly approaches $\phi_{RLP}$ when plotted against the
dimensionless group $R$. However, unlike the plot of $\phi$ vs. $S$,
the asymptotic limit is attained at different values of $R$ for
different materials, leading us to prefer $S$ as the best candidate
for the relevant control parameter.

Additionally, previous measurements of collisions of fluid-immersed
spheres on surfaces\cite{Gondret2002} also show that the Stokes number $St$ -- and not
Reynolds number $Re$ -- is the relevant dimensionless parameter that
defines the onset of bounce-free collisions.  In
addition to supporting the physical picture that the RLP limit
corresponds to conditions where sequentially added spheres settle in
the first mechanically stable location that they encounter, there is
good quantitative agreement between the scale of $St$ where bouncing
ceases, and where $\phi_{RLP}$ is attained.  This correspondence is
shown in Fig.~\ref{restCoeff}. The Stokes number has also been
identified as the parameter controlling the behavior of avalanches
in fluid-immersed piles\cite{duPont2003}.

We now return to the observation that the two curves in Fig.~\ref{phivsparamsFit}
corresponding to steel and acrylic spheres
approach different values of $\phi_{RLP}$ of $0.551$ and $0.540$
respectively.  With the mass density of the materials already
accounted for, the only relevant differences are those of the
contact mechanics of the spheres. In particular, the effective
coefficient of static friction $\mu_s$ between spheres is different
for these materials. Thus, unlike the RCP limit, the RLP limit is
not a purely geometric problem, but involves the mechanics of
interactions.

In order to more fully explore this observation, we have
prepared packings of five different materials (Table~\ref{beadtable}) in the $St\rightarrow 0$ limit.
The coefficient of static friction, $\mu_s$ is affected both by material as well as
by surface topography; indeed we find that $\phi_{RLP}$ does not show a simple trend with surface roughness.  To directly probe the material property of interest, we have devised a method to measure $\mu_s$ for sphere-on-sphere contacts between spheres immersed in a liquid.  The schematic diagram inset in Fig.~\ref{frictionFit} shows the geometry of the setup:  two spheres sit under gravity in the shallow v-groove of a track in the vertical plane.  As the track rotates slowly, static friction between beads prevents them from rolling so they move with the track until they reach a maximal angle $\theta_{max}$.  At $\theta_{max}$ the tangential force between the spheres exceeds the maximum value allowed by the finite coefficient of static friction $\mu_s$ and the beads roll to a lower angle $\theta_{min}<\theta_{max}$.  $\mu_s$ can be then calculated from $\theta_{max}$\footnote{The rotation rate is set slow enough that viscous and centrifugal forces on the particles are negligible compared to gravity and friction.  In this limit, $\mu_s=((L/r)^2-1)/(L/s+1)\,\tan(\theta_{max})$ where $s=r\cos(\alpha)$.  In our setup, $L=6.35\pm0.03\,mm$, $r=1.59\pm0.03\,mm$ and $\alpha=10.0\pm0.5^{\circ}$.   $L$ is the distance from the center of the track to the centers of the spheres whereas the radius of the track at contact with the beads is $L+s$.}.
In Fig.~\ref{frictionFit} we show $\phi_{RLP}$ decreases monotonically as the measured friction coefficient increases.
The dependence on friction coefficient, $\phi_{RLP}(\mu_s)$, is consistent with
$\phi_{RLP}\rightarrow \phi_{RCP}$ as the coefficient of static
friction, $\mu_s \rightarrow 0$.

Our measurements of $\mu_s$ are made under conditions similar to those
in our packing experiments, under the same small loads and fluid
environments, and with sphere-on-sphere contacts that allow for both sliding
and rolling.  We are not aware of any other measurements of interparticle
friction in this regime.  The values of $\mu_s$ we observe are larger than those from our everyday experience at larger normal loads.  Our data thus suggest a resolution of the puzzle that friction coefficients in simulated contact mechanics models\cite{Zhang2001,Silbert2002,Song2008} were thought to be surprisingly large in order to achieve volume fractions as low as seen in experiments.

Finally, we turn from the $RLP$ limit to the large Stokes regime.
Surprisingly, we see a plateau in the value of $\phi$, well
separated from the low Stokes limit. The plateau $\phi$ does not
scale simply with friction coefficient $\mu_s$, and presumably also
involves other particle properties such as inelasticity. We speculate that this plateau value may be related to the ``critical state'' of soil mechanics.
The transition from this plateau to $\phi_{RCP}$ is clearly of
interest, but is not easily explored by merely tuning $S$ with fluid
parameters.

\section{Conclusion}

The strengths of our experiments are (i) the extremely
well-controlled sphericity and monodispersity of the particles, (ii)
detailed characterizations of relevant particles properties such as
surface roughness and of friction coefficient under deposition
conditions, (iii), broad coverage of fluid parameters, and (iv)
employing large enough particles to be well outside the influence of
any attractive interactions. Thus our experiments provide the best available
approximation to the idealized problem of the packing of monodisperse
frictional spheres.  Despite our choice of experimental
geometry, the relatively small system size in our experiments
may introduce wall-effects.  It is clear, however,  that the values of
$\phi_{RLP}$ obtained by the sequential deposition in our experiment
are comparable with those obtained by collective
packing schemes\cite{Onoda1990, Jerkins2008}, and therefore our findings should also
be applicable to those packing protocols.

Previous experiments have shown that looser packings
may be prepared if sedimentation is done more gently.
Our results strengthen this intuitive expectation in three significant directions.
The first is that we arrive at a sharp definition of ``gentle'' deposition:
this limit is governed by the Stokes number,
$St$. The second conclusion is that the RLP limit is
achieved at a nonzero threshold value of $St$,  below which particles entering the packing do not have the ability to explore the landscape of
deposited particles, or to rearrange it. 
Finally, since we eliminate the effect of attractive forces by packing at finite values of $g_f$, we establish the existence of a $\phi_{RLP}$ for cohesionless spheres.

We also provide the first direct experimental study of the
friction-dependence of $\phi_{RLP}$  by measuring the friction
coefficients between particles at small normal loads. Friction
stabilizes packings at volume fractions considerably below $\phi_{RCP}$.
The evolution of a packing from the RLP boundary to the RCP boundary,
and the structure and mechanical properties of the intermediate
states remain largely unexplored.

\section{Acknowledgments}
We acknowledge funding from the NSF through NSF-DMR 0606216 and 0907245,
and the use of facilities funded by NSF-MRSEC 0820506. We thank C. S. O'Hern for valuable conversations.

\bibliographystyle{rsc}
\bibliography{bibliography}

\end{document}